\documentclass[a4paper,11pt,reprint,onecolumn,longbibliography,eqsecnum,floats,aps,amsmath,amssymb,nofootinbib,prd,notitlepage]{revtex4-1}
 \usepackage[a4paper,top=3cm,bottom=3cm,left=2cm,right=2cm,marginparwidth=1.75cm]{geometry}
\usepackage[toc,page]{appendix}
\usepackage[english]{babel}
\usepackage[utf8]{inputenc}
\usepackage{amsmath}
\usepackage{amssymb}
\usepackage{lmodern}
\usepackage{color}
\usepackage{nccmath}
\usepackage{graphicx}
\usepackage{wrapfig}
\usepackage{enumerate}
\usepackage{hyperref}
\usepackage{braket}
\usepackage{float}
\usepackage{dsfont}
\usepackage{mathrsfs}  
\usepackage{natbib}

\newcommand{\rd}{{\rm d}}

\usepackage{indentfirst}

\DeclareMathAlphabet{\dutchcal}{U}{dutchcal}{m}{n}
\SetMathAlphabet{\dutchcal}{bold}{U}{dutchcal}{b}{n}
\DeclareMathAlphabet{\dutchbcal} {U}{dutchcal}{b}{n}
\begin{document}

\title{Quantum dust collapse with cosmological constant and methods for constructing conformal diagrams}

\author{Micha{\l} Bobula}
	\email{michal.bobula@uwr.edu.pl}
	\affiliation{Institute for Theoretical Physics, Faculty of Physics and Astronomy, University of Wroc{\l}aw, pl. M. Borna 9, 50-204  Wroc{\l}aw, Poland}
\author{Tomasz Paw{\l}owski}
	\email{tomasz.pawlowski@uwr.edu.pl}
	\affiliation{Institute for Theoretical Physics, Faculty of Physics and Astronomy, University of Wroc{\l}aw, pl. M. Borna 9, 50-204  Wroc{\l}aw, Poland}

\begin{abstract}
    The loop quantum gravitational collapse of the dust ball in presence of positive cosmological constant is investigated within the Oppenheimer-Snyder collapse scenario. The dust ball interior is described within the framework of loop quantum cosmology, while its exterior geometry is determined by the differentiability of the spacetime metric at the dust ball surface and the assumption of the (vacuum) exterior to be stationary. In order to determine the global causal structure of the investigated spacetime a robust (numerical) method of constructing Penrose-Carter diagrams is built. Unfortunately the presence of cosmological constant does not cure the problems already present in the case of it vanishing --- the exterior geometry resembles that of Reissner-Nordstr\"om-de Sitter black hole, in particular featuring timelike singularties.
\end{abstract}

\maketitle
\section{Motivation}
The problem of the fate of collapsing matter remains one of the most mysterious problems challenging the validity of a physical theory. The classical theory of gravity, General Relativity (GR), while successful when applied to various astrophysical scenarios, fails (at the theoretical level) to provide a fully satisfactory description of the gravitational collapse. The reason behind this fact is the existence of singularities in the theory. They not only indicate a breakdown of a theory itself but also are responsible for the so-called black hole information paradox, a puzzle arising in the framework of quantum fields propagating on black hole backgrounds. For the abovementioned reasons, one has to search for alternatives of GR to model the gravitational collapse. In particular, one can ask whether a quantum description of gravity could resolve the problem. A promising paradigm shift within such a description, the so-called \emph{black-to-white hole transition}, could provide physical picture resolving paradoxes of the (quantum) gravitational collapse \cite{Stephens:1993an, Ashtekar:2005cj, Haggard:2015iya, Bianchi:2018mml, Martin-Dussaud:2019wqc, DAmbrosio:2020mut}. 

The framework of Loop Quantum Gravity (LQG), more precisely its symmetry-reduced variants, seem to provide models realising that new paradigm \cite{BobulaLOOPS, Bobula:2023kbo, Lewandowski:2022zce, Han:2023wxg, Giesel:2022rxi, Giesel:2023tsj, Giesel:2023hys, Giesel:2024mps, Kelly:2020lec, Kelly:2020uwj, Husain:2021ojz, Husain:2022gwp, Hergott:2022hjm, Fazzini:2023scu,  Fazzini:2023ova, Cipriani:2024nhx, Wilson-Ewing:2024uad, Wilson-Ewing:2024fxo, Munch:2021oqn, Alonso-Bardaji:2023qgu}. In particular, recently developed quantum-corrected Oppenheimer-Snyder collapse scenario based on Loop Quantum Cosmology (LQC) shifts classical (coming from GR) Schwarzschild-like description to the (quantum) Reissner-Nordstr\"om-like one, where the black hole ``tunnels''\footnote{Here the analogy with quantum tunnelling is partial only: while the origin of transition is quantum and the observers must pass through the quantum effects dominated region in order to pass to the white hole region, the quantum geometry remains semiclassical for sufficiently massive black holes.} into a white hole \cite{BobulaLOOPS,Bobula:2023kbo, Lewandowski:2022zce, Fazzini:2023scu}. However, the presence of a timelike singularity challenges the reliability of the model used. One could expect, that this deficiency is cured in more realistic, richer ones. For that reason, here we consider the expansion of that model, admitting positive cosmological constant and the fate of the timelike singularity within. Although such a model was already investigated in \cite{Shao:2023qlt}, here we consider a more robust approach with a well-defined quantum Hamiltonian and subsequently analyze its properties, e.g. the causal structure, rigorously. 

Specifically, we will model the collapsing dust ball in the quantum-corrected Oppenheimer-Snyder scenario, where the dust interior geometry is determined via Loop Quantum Cosmology (LQC) \cite{Bojowald:2005epg, Ashtekar:2011ni, Agullo:2016tjh} framework (see for example \cite{Husain:2011tm} for the genuine quantum description of the relevant scenario). To determine the geometry of its vacuum exterior we will employ the procedure of \cite{Bobula:2023kbo}. Instead of quantising the exterior, the classical (though not necessarily adhering to GR dynamics) exterior geometry will be matched to an effective semiclassical interior one via $(i)$ junction conditions at the dust ball surface --- continuity of the (induced) surface metric and exterior curvature, and, $(ii)$ the presence of the 'stationary' Killing field analogous to the time translation one admitted by Schwarzschild-DeSitter metric. As in the vanishing cosmological constant case, these two conditions allow us to determine the exterior metric by the interior one uniquely. Nevertheless, unlike in previous works on quantum Oppenheimer-Snyder collapse (no cosmological constant) \cite{Bobula:2023kbo, Lewandowski:2022zce}, initially we will work with the exterior being time-dependent and only after performing the related analysis of the junction conditions we will conclude that the stationarity of the exterior (existence of the timelike Killing vector) is a necessary assumption of the model. Indeed, the presented analysis can serve as possible route for exploring possible time-dependent exterior geometries.

Another objective of this work is to present a refinement of methods for the construction of conformal (Penrose-Carter) diagrams. In the majority of the works within the field, the spacetime global causal structure is determined by finding the similarities in the form of the metric with known exact solutions, for which that structure is known. However, as physical scenarios and models of higher levels of complication are investigated, such an approach becomes insufficient, especially in cases, when the spacetime metric has to be determined numerically. In such a situation one needs a reliable systematic method of probing the causal structure by numerical means. We present one such method, based on ray-tracing (analysis of null geodesics) refined for general spherically-symmetric spacetimes, and specifically tailored to those admitting a distinguished timelike tube (serving as a point of departure for propagating null geodesics). Such a method allows us to present the spacetimes built by gluing the geometries in two regions to be represented in a single coordinate chart. The presented method will then be used to analyze the collapse scenario introduced a few paragraphs earlier (which then will also serve as an example of the application of the method).

The article is structured as follows: First, in Sec. \ref{secLQC} we recall the genuine quantum description of the dust ball interior in the LQC framework, providing in particular the effective trajectory of the dust ball size in dust emergent time. Next, in Sec. \ref{secEXT} the construction of the exterior of the dust ball is performed via methods already presented in \cite{Bobula:2023kbo}. In particular, the gluing conditions in the absence and presence of stationarity are discussed.  
Subsequently, in sec~\ref{secMethod} we present a numerics-based systematic method of building the Penrose-Carter diagrams devised for spherically symmetric spacetimes constructed by gluing two regions along a timelike tube. The algorithm of reproducing the spacetime's global causal structure is introduced for the general case, without assuming additional symmetries besides the spherical one.
The devised method is finally used in Sec.~\ref{secDIAG} to analyse the Oppenheimer-Snyder model of the collapsing dust in the presence of the positive cosmological constant. The main results of this investigation and their consequences are then summarized in the concluding Sec.~\ref{secConclusions}.

\section{Dust Loop Quantum Cosmology with Cosmological Constant} 
\label{secLQC}

Let us first recall (and expand upon) the treatment of the interior of the collapsing dust ball. As mentioned earlier, we restrict our consideration to the Oppenheimer-Snyder collapse scenario, where the geometry (and matter) admit foliation by homogeneity surfaces. In that case, the matter ball interior can be represented as a region in the flat isotropic Friedman-Robertson-Walker (FRW) universe. For this description, we employ the LQC framework describing the isotropic geometry coupled to (irrotational) dust matter \cite{Husain:2011tk} following the well-established hybrid (polymer quantization for geometry, Schr\"odinger one for matter) quantization procedure of \cite{Bojowald:2008zzb, Ashtekar:2006wn} The dust field $T$ serves here as internal clock, allowing to cast the (sole nontrivial) Hamiltonian constraint as an evolution equation analogous to Schr\"odinger one \cite{Husain:2011tm}
\begin{equation} \label{schrod}
i \hbar \partial_T \Psi(\dutchcal{v}, T)=\hat{H}_{\text {grav }} \Psi(\dutchcal{v}, T) \, ,
\end{equation}
acting on states belonging to the (polymer) geometry Hilbert space --- mathematically the space of the square-integrable functions on the Bohr compactification of the real line
\begin{equation}
    \mathcal{H}_{{\rm grav}} = L^2(\bar{\mathbb{R}},\rd\mu_{{\rm Bohr}}) \ .
\end{equation}
evolving in dust time $T$. The space $\mathcal{H}_{{\rm grav}}$ attains in that picture the role of the physical Hilbert space.
The wave function $\Psi(\dutchcal{v}, T)$ (physical state), ``at the moment $T$'', is decomposed with respect to eigenvectors $|\dutchcal{v}\rangle$ of the oriented volume operator $\hat{\dutchcal{v}}$ (proportional to a power of one of the fundamental operators in loop quantization procedure -- the densitized triad flux across unit square). Together with the $U(1)$ holonomy component $\hat{\mathcal{N}}:=\widehat{\exp (i \dutchcal{b} / 2)}$ it forms the set of fundamental geometry operators. They satisfy commutation relation $[\hat{\dutchcal{v}}, \hat{\mathcal{N}}]=-\hat{\mathcal{N}}$ and act on the basis states as multiplication $\hat{\dutchcal{v}} | \dutchcal{v} \rangle = \dutchcal{v} | \dutchcal{v} \rangle $ and shift $\hat{\mathcal{N}}|\dutchcal{v}\rangle=|\dutchcal{v}+1\rangle$. 
The Thiemann-regularized \cite{Thiemann:1997rv, Ashtekar:2006wn} Hamiltonian takes the form \cite{Husain:2011tm}
\begin{equation}
    \hat{H}_{\text {grav }} = -\frac{3 \pi G \hbar^2}{2 \alpha_o} \sqrt{|\hat{\dutchcal{v}}|} \sin ^2(\hat{\dutchcal{b}}) \sqrt{|\hat{\dutchcal{v}}|} + \frac{3\hbar^2}{16\rho_c\alpha_o}\Lambda |\dutchcal{\hat{v}}| \, ,
\end{equation}
where $\Lambda$ is a cosmological constant, $\alpha_o:=2 \pi \gamma \sqrt{\Delta} \ell_{\mathrm{Pl}}^2$ ($\Delta:=4 \sqrt{3} \pi \gamma \ell_{\mathrm{Pl}}^2$ is so-called LQC \emph{area gap}, $\gamma =  0.2375 \dots$ is the Barbero-Immirzi parameter for which we take value as determined in \cite{Domagala:2004jt, Meissner:2004ju}), $\rho_c \approx 0.82 \,  \rho_{\mathrm{Pl}}$ is the critical density of LQC \cite{Ashtekar:2006wn}, and $\sin (\hat{\dutchcal{b}}):=-i / 2\left(\hat{\mathcal{N}}^2-\hat{\mathcal{N}}^{-2}\right)$. 

While the system under consideration has already been described on the genuine quantum level \cite{Husain:2011tm}, its particular application here (gluing to \emph{classical} exterior metric) requires us to reduce our description to just quantum trajectories (of the expectation values). Furthermore, by its very nature, the consistency of the gluing requires us to restrict our studies to very sharply localized semiclassical states. Therefore, for the purpose of this work, it is sufficient to employ the so-called \emph{effective dynamics} \cite{Singh:2005xg} (which in turn is the $0$th order cutoff of the more systematic semiclassical treatment \cite{Bojowald:2005cw}). In this we follow exactly the treatment of \cite{Bobula:2023kbo}. First, we approximate classical treatment by replacing the component operators with their expectation values, assuming further that the (expectation values of the) holonomy components can be represented as exponentiated momentum $\dutchcal{b}$, which gives $\dutchcal{v}:= \langle \hat{\dutchcal{v} } \rangle $ and $\sin \dutchcal{b} = \langle(i / 2)(\hat{\mathcal{N}}^2-\hat{\mathcal{N}}^{-2})\rangle$, $ \cos \dutchcal{b} =  \langle(1 / 2)(\hat{\mathcal{N}}^2+\hat{\mathcal{N}}^{-2})\rangle)$. The chosen variables form a Poisson algebra with the bracket $\{v, b\}=-2 / \hbar$ and the effective Hamiltonian, built by replacing all the component operators with listed variables/functions of $\dutchcal{b}$, takes the form
\begin{equation}
    H_{\mathrm{eff}}(\dutchcal{v}, \dutchcal{b}) = -\frac{3 \pi G \hbar^2}{2 \alpha_o} \dutchcal{v} \sin^2 (\dutchcal{b}) +  \frac{3\hbar^2}{16\rho_c\alpha_o} \Lambda \dutchcal{v} \ .
\end{equation}
The evolution of the resulting effective system is then determined by the set of Hamilton-Jacobi equations
    \begin{equation}
    \left\{  
    \begin{aligned}
    \dot{\dutchcal{v} } &= \{\dutchcal{v}, H_\mathrm{eff}\} = - 2 A \sin (2 \dutchcal{b})  \, ,\\
    \dot{\dutchcal{b} } &= \{\dutchcal{b}, H_\mathrm{eff}\} = - 2 A \sin^2 \dutchcal{b} + 2 C \, . \\
    \end{aligned}
    \right.
\end{equation}
where $A:=\frac{3 \pi G \hbar}{2 \alpha_o}$, $C:=  \frac{3\hbar}{16\rho_c\alpha_o} \Lambda$, and the dot indicates differentiation w.r.t. $T$. Upon setting the initial conditions $\dutchcal{v}(T=0) = \dutchcal{v}_\mathrm{in} $ and $\dutchcal{b}(T=0) = \pi/2$ we fix the time translation freedom, setting the moment $T=0$ to correspond to the \emph{bounce}, with $\dutchcal{v}_\mathrm{in}$ being the minimal (bounce) volume of a particular trajectory. Solving this initial value problem and using the relation between the scale factor $a(t)$ and the variable $\dutchcal{v}$: $a(T) = V^{1/3} = (\alpha_o \dutchcal{v}(T))^{1/3}$, we can describe the dust ball interior by the effective metric
\begin{equation} \label{frw}
\mathrm{d} s_{-}^2=g^-_{\alpha \beta} \mathrm{d} x^\alpha \mathrm{d} x^\beta=-\mathrm{d} T^2+a(T)^2 \mathrm{~d} r^2+r^2 a(T)^2 \left( \mathrm{d} \theta^2 + \sin ^2 \theta \mathrm{d} \varphi^2 \right)
\end{equation}
where the scale factor is of the form
\begin{equation} \label{scaleanalytic}
    a(T) =   \left(  \frac{\alpha_o \dutchcal{v}_\mathrm{in}}{2 C} \left(A \cosh \left(4 T \sqrt{C
   (A-C) } \right)-A+2 C\right) \right)^{1/3}
\end{equation}
This effective metric is defined for all $\Lambda\in (0,\Lambda_c:= 8\pi G \rho_c)$ and all positive $\dutchcal{v}_{\rm in}$, however, it approximates the quantum trajectory with decent accuracy only when $\dutchcal{v}_{\rm in}\gg 1$.
The inherent (for the metrics of the form \eqref{frw}) freedom of rescaling the comoving radial coordinate $r$ is further fixed by setting the ball comoving radius (as in \cite{Bobula:2023kbo}) to the one of unit volume ball: $r=r_b=(4 \pi / 3)^{-1 / 3}$ .

\section{Exterior metric} 
\label{secEXT}

Outside of the dust ball we are going to assume the lack of matter, though (as we consider a quantum modification to GR) we cannot assume that the vacuum Einstein field equations still hold. Instead, the (otherwise general spherically symmetric) geometry will be restricted by purely geometric properties. The first of them, natural from the physical point of view (non-singular transition from the dust ball to its exterior) is that the metric is of $C^1$ class at the dust ball surface $\Sigma$, which in our description forms a boundary, on which two distinct geometries are being glued. This condition is mathematically equivalent to a set of junction conditions. First, we will discuss these conditions below in the context of gluing the exterior to the FRW geometry described by metric \eqref{frw} (with the scale factor considered for the time being to be an arbitrary function of time). They will be applied further in Sec.~\ref{extsubSEC} to derive the explicit form of the metric for the case of scale factor corresponding to the LQC geometry \eqref{scaleanalytic} determined in the previous section.

To begin with, we assume that the following line element describes the exterior region
\begin{equation} \label{ext}
    \mathrm{d}s^2_+ = g^+_{\alpha \beta} \mathrm{d} x^\alpha \mathrm{d} x^\beta = -F(t,X) \mathrm{d}t^2 + G(t,X) \mathrm{d} X^2 + X^2 \left( \mathrm{d} \theta^2 + \sin ^2 \theta \mathrm{d} \varphi^2 \right)
\end{equation}
with yet-to-be-determined (and so far independent) functions $F(t,X)$ and $G(t,X)$. The $\Sigma$ as seen from the interior can be parametrized with the dust internal time $T$ as $x^\alpha = (T, r=r_b = \mathrm{const.} )$, whereas for the proposed exterior description we have $x^\alpha = (t=\dutchcal{t}(T), X=R(T) )$ respectively. We further introduce on $\Sigma$ a convenient auxiliary structure -- a dyad of vectors $l^\alpha, n^\alpha$ adapted to spherical symmetry such that (i) $l^\alpha$ is the four-velocity of an observer comoving with $\Sigma$ (being by construction a timelike tube), that is $l^\alpha = \partial x^\alpha / \partial T $ and (ii) $n^a$ is a normalized vector normal to the surface $\Sigma$, that is satisfying the relations $l^\alpha n_\alpha =0$ $n^\alpha n_\alpha =1$. 
The condition for the metric to be $C^1$ at the boundary surface takes the form of the Israel-Darmois junction conditions imposed on its interior and exterior forms (\ref{frw}, \ref{ext}). To write them down explicitly, let $y^a = (T, \theta, \varphi)$ be a coordinate system at $\Sigma$, and $ \textbf{h}^+ := \textbf{g}^+\big|_\Sigma  $, $\textbf{h}^- := \textbf{g}^-\big|_\Sigma  $. Then, these conditions form a set consisting of $a)$ continuity of the induced metrics $h^+_{ab} =h^-_{ab} $, and $b)$ continuity of extrinsic curvatures $K^+_{ab} =K^-_{ab} $, where $K_{a b}:=n_{\alpha ; \beta} \frac{\partial x^\alpha}{\partial y^a} \frac{\partial x^\beta}{\partial y^b}$ (see Chapters 3.7-3.8 of \cite{Poisson:2009pwt} for a complete discussion). These matching conditions form a soluble system of algebraic equations. In particular, from $h_{\theta \theta}^+ =h^-_{\theta \theta} $ one gets 
\begin{equation} \label{Rsurf}
   X\Big|_\Sigma = R(T) = r_b a(T) \, .
\end{equation}
Subsequently, equations $h^+_{TT}=h^-_{TT}$, $K^+_{\theta \theta} = K^-_{\theta \theta}$, and \eqref{Rsurf} combined together deliver 
\begin{equation} \label{GT}
    G = \frac{1}{1-R'(T)^2} \, ,
\end{equation}
and 
\begin{equation} \label{gf}
    \dutchcal{t}'(T)^2 = \frac{G}{F} \, .
\end{equation}
Finally, the condition $K_{T T}^{-}=K_{T T}^{+}$ gives
\begin{equation} \label{TT}
    n_{\alpha ; \beta} l^\alpha l^\beta=n_{\alpha} l^\alpha_{;\beta} l^\beta =0 \, .
\end{equation}
At this point we note, that due to a physical role of $\Sigma$ (as the surface formed by trajectories of free-falling dust particles) a condition stronger than \eqref{TT} must be satisfied. Since the four-acceleration $l^\alpha_{;\beta} l^\beta$ identically vanishes as seen from the interior, we demand the same as seen from the exterior (so that the covariance is preserved). With this condition, \eqref{TT} is automatically fulfilled. Explicitly, the following holds (where $l^\alpha = (\dutchcal{t}'(T), R'(T) )$
\begin{equation} \label{lt}
     l^t_{;\beta} l^\beta = 0 \, ,
\end{equation}
and 
\begin{equation} \label{lX}
     l^X l_{;\beta} l^\beta = 0 \, .
\end{equation}
The relation \eqref{lt} combined together with \eqref{GT}, \eqref{gf}, and with the decomposition of the derivatives of $F,G$ on $\Sigma$
\begin{subequations} \label{FpGp}
    \begin{align}
        F'(T) &= \partial_t F(t,X) \Big|_\Sigma \frac{\mathrm{d} t}{\mathrm{d} T } +  \partial_X F(t,X) \Big|_\Sigma \frac{\mathrm{d} X}{\mathrm{d} T }\\
        G'(T) &= \partial_t G(t,X) \Big|_\Sigma \frac{\mathrm{d} t}{\mathrm{d} T } +  \partial_X G(t,X) \Big|_\Sigma \frac{\mathrm{d} X}{\mathrm{d} T }
    \end{align}
\end{subequations}
deliver\footnote{Keep in mind that we can write $\frac{\mathrm{d} t}{\mathrm{d} T } = \dutchcal{t}'(T) $, and  $\frac{\mathrm{d} X}{\mathrm{d} T } = R'(T)$.}
\begin{equation} \label{lastjunction}
\begin{split}
   l^t_{;\beta} l^\beta &= \dutchcal{t} ''(T) + \frac{R'(T) \dutchcal{t} '(T) \partial_X F(t,X) \Big|_\Sigma}{F}+\frac{\dutchcal{t} '(T)^2 \partial_t F(t,X) \Big|_\Sigma }{2 F}+\frac{R'(T)^2 \partial_t G(t,X) \Big|_\Sigma }{2 F}=  \\
   &\pm \sqrt{G-1} \sqrt{G/F} \, (G \, \partial_X F(t,X) \Big|_\Sigma + F \, \partial_X G(t,X) \Big|_\Sigma  ) + (2 G -1) \sqrt{G} \partial_t G(t,X) \Big|_\Sigma =0 \, .
\end{split}
\end{equation}
In the above, the $\pm$ sign indicates positive or negative solution to \eqref{GT}, that is, $R'(T) = \pm \sqrt{1-G}/\sqrt{G}$. The remaining condition \eqref{lX} is automatically satisfied when combined together with the \eqref{lastjunction}, \eqref{GT}, \eqref{gf} and \eqref{FpGp}. 

So far, we completely characterized junction conditions following from the requirement of a nonsingular transition through the dust ball surface. As we can see, they are relatively mild and without further assumptions or imposing field equations do not suffice to uniquely determine the exterior metric by the interior one. Even in the case when the exterior satisfied the vacuum Einstein field equations (which we cannot assume here), the boundary data at $\Sigma$ (being a timelike tube) would not suffice to determine the exterior geometry. One can however hope to further restrict the latter by imposing additional conditions.

In GR, when considering matter collapse in otherwise vacuum universe, it is natural to expect that the vacuum exterior of the collapsing matter remains stationary. We will also consider this condition further in the paper, however before imposing it,
as an aside, let us consider 2 cases where we impose some weaker assumptions. These assumptions (together with their consequences) are listed as follows.
\begin{enumerate}[(i)]
  \item First, suppose that $G(t,X) \rightarrow G(X)$ on the whole exterior. Such condition implies in particular via \eqref{lastjunction}, that at the boundary we have $\partial_X (FG)\Big|_\Sigma =0$, thus $FG=constant$, which without the loss of generality can be set to $FG=1$. This in turn implies the relation $F(X) = 1/G(X)$ at $\Sigma$. \label{it:assumption1}
  \item Alternatively, we can start with the global assumption $F(t,X)=1/G(t,X)$. Then \eqref{lastjunction} implies the condition $\partial_t G(t,X) \Big|_\Sigma = 0 $.
\end{enumerate}
Neither the above assumptions, nor their implications are sufficient for ensuring global time-independence of every metric component of \eqref{ext} -- they do not imply the stationarity of the exterior metric. They are too weak to determine the exterior of the metric uniquely. In order to ensure the latter we need to assume the stationarity of the metric \eqref{ext} explicitly.
Thus, throughout the rest of the work we assume the existence of the timelike Killing vector in the exterior, that is, we incorporate $F(X,t) \rightarrow F(X) = 1/G(X)$ and $G(X,t) \rightarrow G(X)$. We also require that $F(X)=1/G(X)$, which under the assumptions made can be done without the loss of generality (see the discussion of the point \eqref{it:assumption1} above).

\subsection{Explicit exterior metric for the dust loop quantum gravitational collapse with cosmological constant.} \label{extsubSEC}

Let us study the specific consequences of the conditions discussed above on the exact form of the exterior geometry, given that the interior one is determined by the scale factor given by \eqref{scaleanalytic}\footnote{Note that the time-reflection symmetry of the scale factor allows us to write $R'(T) = \mathrm{sgn} (T) \frac{\sqrt{G-1}}{\sqrt{G}}$, because $R'(T) >0$ for $T>0$ and $R'(T) <0$ for $T<0$. Due to this fact, we can exchange the $\pm$ sign in \eqref{lastjunction} with $\mathrm{sgn}(T)$.}. To write $F$ and $G$ as functions of $X$ coordinate, we invert the relation \eqref{Rsurf} to get 
\begin{equation} \label{TS}
    T(X) = - \frac{1}{4 \sqrt{C(C-A)} } \arccos \left( 1 + \frac{2 C}{A} \left( \frac{X^3}{\alpha_o  r_b^3 \dutchcal{v}_{\mathrm{in}}} -1\right)\right) \, .
\end{equation}
Note that now we can plug \eqref{TS} to both \eqref{GT} and $F(T) = 1-R'(T)^2$, that is,  $F(X) = 1/G(X) = 1-R'(T(X))^2$. The formulae not only hold at $\Sigma$, but also in the whole exterior -- the ``stationarity '' (existence of a Killing field $\partial_t$) of metric \eqref{ext} together with the symmetry $T \rightarrow-T$ in \eqref{Rsurf} allow the propagation of $F$ and $G$ through the exterior along surfaces of constant $X$. Ultimately we can write
\begin{equation} \label{FX}
    F(X) = 1/G(X)= 1 - \frac{2 G M_\mathrm{eff}}{X} + \frac{9 G^2 M_\mathrm{eff}^2 }{4 \left(A-2 C \right)^2 X^4  } - \frac{16}{9} \left( A - C \right) C X^2 \ ,
\end{equation}
where $M_{\mathrm{eff}}:=8 \alpha_o  r_b^3 \dutchcal{v}_{\mathrm{in}} (A-C)(A-2C)/ (9 G)$. As an aside, the metric \eqref{FX} was also found in \cite{Lin:2024flv} by solving loop quantum effective equations of motion in spherical symmetry accounting for the presence of a cosmological constant in midisuperspace approach. This result further supports the viability of the junction conditions formalism we employed.
Throughout the rest of our work, we will restrict our analysis to the situation when \eqref{FX} has 3 positive roots, say $X_1 < X_2 < X_3$ --- this corresponds to the situation when $M_\mathrm{eff} \gg m_\mathrm{Pl}$ is comparable to the mass of collapsing matter in astrophysical scenarios and $\Lambda$ is close to observational values --- see Appendix \ref{approots} for the complete discussion. In other words, we consider a dust ball collapse ``slightly perturbed''\footnote{This is not the perturbation in the sense that the system is linearized. We simply look for the dynamics when $\Lambda$ is \emph{small}. } by the presence of the cosmological constant. For notational purposes we also define $T_1 := T(X_1)$, $T_2 := T(X_2)$, and $T_3 := T(X_3)$ according to \eqref{TS}. The exterior geometry is completely determined now. The line element \eqref{ext} (with $F(X)$ given in \eqref{FX}) suggest that the exterior describes black-hole-like, asymptotically de Sitter spacetime. We will indeed confirm that statement in the next sections.

\section{Methods for constructing conformal diagrams} 
\label{secMethod}

In this section, we will propose a scheme for (numerical) derivations of \textit{Penrose-Carter diagrams} for a class of spacetimes comprising two componential ones with possibly different coordinate systems. As an example, one can think about two componential spacetimes smoothly glued together. The procedure is going to serve two purposes: $a)$ identifying the positions of the spacetime points on the would-be-generated diagram, and $b)$ extending a coordinate system chosen for one region of the componential spacetime (for example describing the interior region) to the one across the boundary through which the regions are connected (for example describing the exterior). The latter, in other words, allows to cover the whole joint spacetime (or some region of it where two componential spacetimes are included) with one extended coordinate system. The crucial role in this construction will be played by the geometry of the boundary between the two spacetimes. The proposed strategy will be then applied to the spacetime derived in Sections \ref{secLQC} and \ref{secEXT} in the next section.~\ref{secEXT}.

To start with, let's assume we have two spherically-symmetric spacetimes smoothly joined at the timelike boundary surface $\Omega$. Although in this work we assumed the spherical symmetry, though the key steps of the scheme may also work for more general classes of spacetimes. Furthermore, we assume that the metric tensors are Lorentzian and nondegenerate. Consequently we can rewrite the metrics, say $\mathbf{g}^{\mathrm{in}}$ and $\mathbf{g}^{\mathrm{out}}$, in double-null coordinates compatible with the spherical symmetry. That is, let $(u_*^{\mathrm{in}}, v_*^{\mathrm{in}} )$ be a pair of the null coordinates compatible with spherical symmetry for the first componential spacetime, and $(u_*^{\mathrm{out}}, v_*^{\mathrm{out}} )$ be the analogous one for the second one respectively. We can then write respective line elements as
\begin{equation} \label{linein}
    g^{\mathrm{in}}_{\mu\nu} \mathrm{d} x^{\mu} \mathrm{d }x^{\nu} = -\mathcal{F}^{\mathrm{in}}(u_*^{\mathrm{in}}, v_*^{\mathrm{in}}) \mathrm{d} u_*^{\mathrm{in}}  \mathrm{d} v_*^{\mathrm{in}} + \dots \, ,
\end{equation}
and
\begin{equation} \label{linout}
    g^{\mathrm{out}}_{\mu\nu} \mathrm{d} x^{\mu} \mathrm{d }x^{\nu} = -\mathcal{F}^{\mathrm{out}}(u_*^{\mathrm{out}}, v_*^{\mathrm{out}}) \mathrm{d} u_*^{\mathrm{out}}  \mathrm{d} v_*^{\mathrm{out}} + \dots \, ,
\end{equation}
where the dots indicate the angular parts of the line elements. We proceed with the following steps
\begin{figure}[h] 
    \centering
    \includegraphics[width=7cm]{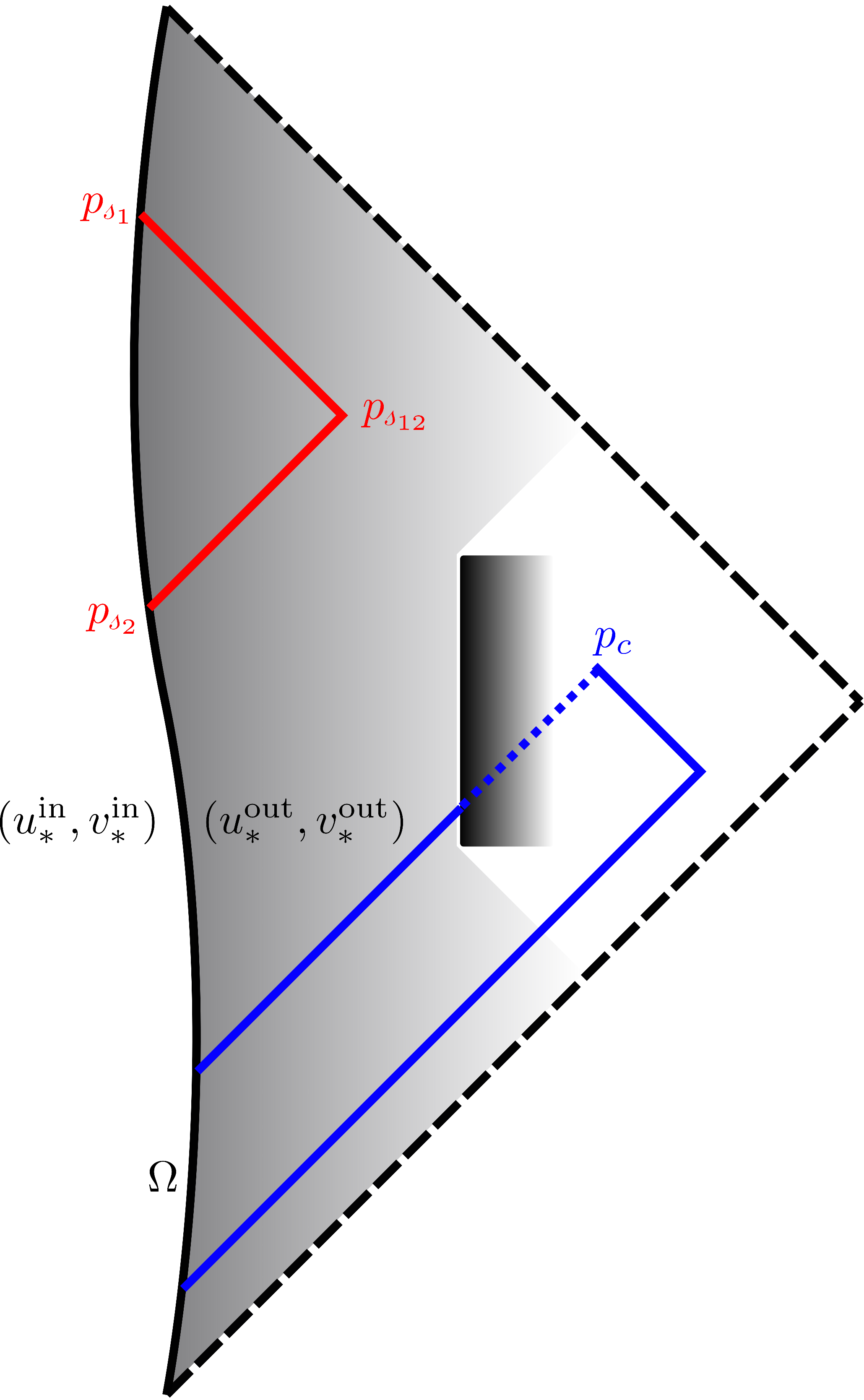}
    \caption{Extension of a coordinate system visualized. Region on the left-hand (right-hand) side from the timelike boundary surface, that is wavy line denoted with $\Omega$, is covered with the pair $(u_*^{\mathrm{in}}, v_*^{\mathrm{in}})$ (with the pair $(u_*^{\mathrm{out}}, v_*^{\mathrm{out}})$). Let $\dutchcal{s}$ be the parameter at $\Omega$ -- the trajectory of the boundary surface is given either by $(u_*^{\mathrm{in}}(\dutchcal{s}), v_*^{\mathrm{in}}(\dutchcal{s}))$ or $(u_*^{\mathrm{out}}(\dutchcal{s}), v_*^{\mathrm{out}}(\dutchcal{s}))$. The point $p_{\dutchcal{s}_{12}}$ in the extended coordinate system \eqref{extending} is given by  $ ( u_*^{\mathrm{in}} ( \dutchcal{s} (u_*^{\mathrm{out}} \big|_{p_{\dutchcal{s}_{12}  }} )  ), v_*^{\mathrm{in}} ( \dutchcal{s} (v_*^{\mathrm{out}} \big|_{p_{\dutchcal{s}_{12}  }} )  ) ) =  (u_*^{\mathrm{in}} ( \dutchcal{s}_2),  v_*^{\mathrm{in}} ( \dutchcal{s}_1  ))$. Dashed lines indicate the last light ray reaching $\Omega$ and the first ray leaving it. Due to the construction, they enclose the region covered by \eqref{extending}. Dark rectangle represent a possible impassable barrier for light rays (for example a timelike singularity) or just a region not representing any point of the spacetime. The point $p_c$ cannot be reached by a ``straight'' line (geodesic) from $\Sigma$, however, it can be reached by a zig-zag line being the combination of $v_*^{\mathrm{out}}=const.$ and $u_*^{\mathrm{out}}=const.$ geodesics (equivalently $v_*^{\mathrm{in}} (v_*^{\mathrm{out}}) ) =const. $ and $u_*^{\mathrm{in}} (u_*^{\mathrm{out}}) ) =const. $). }
     \label{uklady}
\end{figure}
\begin{itemize}
    \item Let $\dutchcal{s}$ be a monotonic and compatible with spherical symmetry function at $\Omega$. In other words, $\Omega$ can be seen as a ``tube'' where values of $\dutchcal{s}$ label 2-spheres. For example, the ``parameter'' $T$ from Section \ref{secEXT} is such a function. Based on junction conditions, solve for the ``trajectory'' of $\Omega$ with respect to both coordinate systems, namely, calculate $(u_*^{\mathrm{in}}(\dutchcal{s}), v_*^{\mathrm{in}}(\dutchcal{s}) )$ and $(u_*^{\mathrm{out}}(\dutchcal{s}), v_*^{\mathrm{out}}(\dutchcal{s}) )$. Find inverses, that is $\dutchcal{s}(u_*^{\mathrm{out}})$ and $\dutchcal{s}(v_*^{\mathrm{out}})$. Note that the functions may be only piecewise invertible, however, the construction will still work. These coordinates will provide an abstract chart on the to-be Penrose-Cartan diagram through an assumption of them being dragged (constant) along null geodesics, forming the lines at $45^o$ angle on the diagram.
    \item \emph{Collect those points $(u_*^{\mathrm{in}}, v_*^{\mathrm{in}} )$ and $(u_*^{\mathrm{out}}, v_*^{\mathrm{out}} )$, outside of $\Omega$, that actually represent the} spacetime points. It usually happens that not every point belonging to the coordinate chart actually represents a spacetime point. In particular, if the coordinates have compact ranges they can cover not only spacetime points but also sectors without physical significance (outside of an actual spacetime). For example, a conformal diagram of a maximally extended Schwarzschild solution is typically represented by a ``hexagon'' on an appropriate coordinate chart, and the points outside the hexagon do not represent any spacetime points. Motivated by the fact that the situation can be more complicated in general, we aim to deliver a tool useful for determining the points representing the actual spacetime. Note that the tool works in the same way for both componential spacetimes, though here we will present it only for the componential spacetime covered by $(u_*^{\mathrm{out}}, v_*^{\mathrm{out}} )$.

    We aim to analyse the behaviour of an affine parameter at $u_*^{\mathrm{out}} = const.$ and $v_*^{\mathrm{out}} = const.$ families of null geodesics. We stress that the points where the affine parameter reaches infinite value indicate the spacetime endpoints. However, the spacetime may end in a finite value of the affine parameter while traversing along null geodesic --- for example, a light ray can hit a singularity. Consider the following. Suppose we analyze $v_*^{\mathrm{out}} = const.$ geodesic, and let $u_*^{\mathrm{out}}$ be a parameter. The tangent vector to that geodesic, say $k^{\alpha}$, satisfies $k^{\alpha}_{;\beta}k^\beta = \frac{\mathrm{d} \mathcal{F}^{out}(u_*^{\mathrm{out}} )  }{\mathrm{d} u_*^{\mathrm{out}} } \frac{1}{\mathcal{F}^{\mathrm{out}} (u_*^\mathrm{out}) } k^{\alpha}$. It tells us that $u^{\mathrm{out}}$ is not affine, however, it is related to the one, say $\lambda$, via (see Chapter 1.3 of \cite{Poisson:2009pwt} for the general discussion)
    \begin{equation} \label{affineRelation}
        \frac{\mathrm{d}^2 \lambda(u_*^{\mathrm{out}}) }{\mathrm{d}( u_*^{\mathrm{out}})^2 } = \frac{\mathrm{d} \mathcal{F}^{\mathrm{out}}(u_*^{\mathrm{out}} )  }{\mathrm{d} u_*^{\mathrm{out}} } \frac{1}{\mathcal{F}^{\mathrm{out}} (u_*^\mathrm{out}) } \frac{\mathrm{d}\lambda(u_*^{\mathrm{out}}) }{\mathrm{d} u_*^{\mathrm{out}} }
    \end{equation}
    We stress that the method of consideration will work as long as we can solve the above equation, for which the following requirements need to be satisfied. Firstly, initial $\lambda( u_*^{\mathrm{out}})$ and $\mathrm{d} \lambda(u_*^{\mathrm{out}})/\mathrm{d} u_*^{\mathrm{out}} $ (for example given at $\Omega$) needs to be known. Secondly, we need some differential condition (or constraint) for $\mathcal{F}^{\mathrm{out}} (u_*^\mathrm{out})$ letting us solve the above possibly without knowing beforehand the explicit form of $\mathcal{F}^{\mathrm{out}} (u_*^\mathrm{out})$ -- of course, if $\mathcal{F}^{\mathrm{out}} (u_*^\mathrm{out})$ is explicitly known everywhere where needed, then the requirement is automatically satisfied. Note that in GR, $\mathcal{F}^{\mathrm{out}} (u_*^\mathrm{out})$ can be determined by Einstein Field Equations, however, here we discuss a potentially more general situation where metric components may not be determined in that way (for example, see Section \ref{secEXT}). So, here the minimal requirement is the abovementioned differential condition. Afterwards, we obtain the function $\lambda(u_*^{\mathrm{out}})$ (or $u_*^{\mathrm{out}}(\lambda)$) telling us how the affine parameter behaves on the $v_*^{\mathrm{out}} = const.$ geodesic. That knowledge allows us to explore the possible extendibility of the geodesic. If the geodesic is extendible, one should extend it for example with the use of analytic extension. On the other hand, inextendible geodesics (for a finite value of the affine parameter) possibly indicate the existence of singular points or surfaces. Thus one also needs to study the geometrical properties of these to determine their nature. Finally, one just needs to collect all necessary points $u_*^{out}$ (given $v_*^{\mathrm{out}} = const.$).
    
    The scheme presented above works also (completely analogously) for the case of $u_*^{\mathrm{out}}=const.$ family of null geodesic, where we collect $v_*^{out}$ points. In general, we propose to start from $\Omega$ and then exploit all possible points reachable from there. Probably some points cannot be reached by a "straight'' geodesic on a yet-to-be-generated diagram, rather they could be reached by zig-zags being the combination of $u_*^\mathrm{out} = const.$ and $v_*^\mathrm{out} = const.$ geodesics (-- see Figure \ref{uklady}). To go along such a zig-zag, simply solve the equations for the affine parameter with appropriate initial conditions, where each step is given by either $u_*^{\mathrm{out}}=const.$ or $v_*^{\mathrm{out}}=const.$.  
    
    The above procedure is presented for the second componential spacetime covered by $(u_*^{\mathrm{out}}, v_*^{\mathrm{out}} )$, however, one can proceed in completely analogous in the case of the first componential spacetime covered by $(u_*^{\mathrm{in}}, v_*^{\mathrm{in}} )$.
        
    \item \emph{Extend the coordinates.} To cover the joint spacetime within the single coordinate chart we need to extend a coordinate system viable for the chosen componential spacetime to cover also the other one. We can either extend $(u_*^{\mathrm{in}}, v_*^{\mathrm{in}} )$ to cover also the second componential spacetime or we can extend $(u_*^{\mathrm{in}}, v_*^{\mathrm{in}} )$ to cover also the first componential spacetime. We present the prior variant (the latter one is completely analogous). To begin, $(u_*^{\mathrm{in}}, v_*^{\mathrm{in}} )$ have to have compact ranges (if they are not, one can perform compactification, for example $u_*^{\mathrm{in}} \rightarrow \tanh u_*^{\mathrm{in}}$ ). The extended coordinate system, in the second componential spacetime, is now given by
    \begin{equation} \label{extending}
    \left\{  
    \begin{aligned}
     u_*^{\mathrm{in}} (u_*^{\mathrm{out}}) = u_*^{\mathrm{in}} ( \dutchcal{s} (u_*^{\mathrm{out}})  ) \, ,\\
     v_*^{\mathrm{in}} (v_*^{\mathrm{out}}) = v_*^{\mathrm{in}} ( \dutchcal{s} (v_*^{\mathrm{out}})  ) \, . \\
    \end{aligned}
    \right.
\end{equation} 
    The above construction is visualised in Figure \ref{uklady}. The coordinate chart covering the first componential spacetime via $(u_*^{\mathrm{in}}, v_*^{\mathrm{in}} )$ and the second one via \eqref{extending} can be interpreted as an embedding of the spacetime in $\mathbb{R}^2$ plane. The construction will work as long as  $u_*^{\mathrm{in}} ( \dutchcal{s} )$ and $v_*^{\mathrm{in}} ( \dutchcal{s} )$ remain monotonic functions -- the choice of the possible compactification plays here the crucial role. Note that if the functions are piecewise monotonic, one can (in most cases) get the full monotonicity via suitable coordinate transformation.

    Note that the extension through the relation $u^{\rm in}_*\leftrightarrow \dutchcal{s} \leftrightarrow u^{\rm out}_*$ and $v^{\rm in}_*\leftrightarrow \dutchcal{s} \leftrightarrow v^{\rm out}_*$ is realized for each coordinate separately. As illustrated in Figure \ref{uklady}, given a spacetime point of our interest, namely $p_{\dutchcal{s}_{12}}$ coordinatized by the pair $(u_*^{\mathrm{out}} \big|_{p_{\dutchcal{s}_{12}  }}, v_*^{\mathrm{out}} \big|_{p_{\dutchcal{s}_{12}  }}  )$ and located in the second componential spacetime, there is only one, say "right-moving", $u_*^{\mathrm{out}}=const.$ null geodesic connecting $p_{\dutchcal{s}_{12}}$ with $p_{\dutchcal{s}_{2}}$ at $\Omega$. Therefore, we have unique value $\dutchcal{s} = \dutchcal{s}_2$ (and then $u_*^\mathrm{in}(\dutchcal{s}=\dutchcal{s}_2))$  corresponding to that $u_*^{\mathrm{out}} \big|_{p_{\dutchcal{s}_{12}  }} = u_*^{\mathrm{out}} \big|_{p_{\dutchcal{s}_{2}  }} = const.$ null geodesic. Similarly there is only one, say "left-moving", $v_*^{\mathrm{out}}=const.$ null geodesic connecting $p_{\dutchcal{s}_{12}}$ with $p_{\dutchcal{s}_{1}}$ at $\Omega$. Therefore there is only one $\dutchcal{s}=\dutchcal{s}_1$ (and then $v_*^\mathrm{in}(\dutchcal{s}=\dutchcal{s}_1))$) corresponding to that $v_*^{\mathrm{out}} \big|_{p_{\dutchcal{s}_{12}  }} = v_*^{\mathrm{out}} \big|_{p_{\dutchcal{s}_{1}  }} = const.$ null geodesic. 

    Presented extension \eqref{extending} works directly only within a certain region of spacetime -- a kind of ``flipped light cone'' of the (portion of the) boundary surface, which is indicated in Figure \ref{uklady} by the dashed lines, which may not cover the whole spacetime or even its every region of interest.  Nevertheless, one can further extend the extended coordinate system \eqref{extending} beyond that ``cone''. To go ``up'' from the upper dashed line, one has to have a viable extension of the coordinates $v_*^{\mathrm{in}} (v_*^{\mathrm{out}})$ there. One way to obtain it is the analytic extension. In the case, when only numerical form is $v_*^{\mathrm{in}} (v_*^{\mathrm{out}})$ is available (as it actually happens in the examples presented in the next sections), one can in particular perform a polynomial extrapolation. Analogously, a similar procedure can be incorporated for the extension beyond the lower dashed line.

    \item \emph{Generate the conformal diagram.} Plot surfaces, trajectories etc. of interest -- simply plug collected points, $(u_*^{\mathrm{in}}, v_*^{\mathrm{in}} )$ and $(u_*^{\mathrm{out}}, v_*^{\mathrm{out}} )$ to the extended coordinate chart. In the first componential spacetime, the coordinates are given by $(u_*^{\mathrm{in}}, v_*^{\mathrm{in}} )$. In the second spacetime coordinates are given by \eqref{extending}. As already mentioned, we could also proceed the other way around, that is, we could extend $(u_*^{\mathrm{in}}, v_*^{\mathrm{in}} )$ to cover the first componential spacetime. 

    It is worth noting, that in Section \ref{secDIAG} we will see how, due to the staticity of the exterior, the rather complicated formula \eqref{affineRelation} will reduce to simple integral on the base manifold. We also note that it may happen that the collection of points $(u_*^{\mathrm{in}}, v_*^{\mathrm{in}} )$ is given ``for free'' (with no effort) -- this will be the case of the cosmological interior.
\end{itemize}

\section{Conformal diagrams for dust loop quantum collapse with cosmological constant} 
\label{secDIAG}

Having the method developed in Sec. \ref{secMethod}, we will utilize it to construct conformal diagrams corresponding to the model of gravitational collapse derived in Sec. \ref{secLQC}, \ref{secEXT}. We will proceed in a twofold way and each of the strategies will provide ``different-looking'' diagrams --- compare the diagrams in Fig. \ref{prosty} and \ref{confout}. However, the causal structure will qualitatively be the same in both cases. In the first approach, we will particularly extend the interior coordinate system to cover also the exterior (according to the method). In the second one, we will extend the exterior coordinate system to cover also the interior. 

\subsection{Extension of the interior coordinates} \label{extenstionInteriorSEC}
To begin with, we introduce double null coordinates for the interior defined as

\begin{equation} \label{uvin}   
\begin{aligned}
\tilde{u}^{\text{in}}(T, r) = \tau (T) - r \, ,\\
\tilde{v}^{\text{in}}(T, r) = \tau (T) + r \, ,\\
\end{aligned}
\end{equation}
where $\tau(T) = \int_{0}^{T} 1/a(T') \mathrm{d}T'  $. The line element \eqref{frw} written with these coordinates is
\begin{equation}
    \mathrm{d}s^2_- = - a(T)^2 \mathrm{d}u^{\text{in}} \mathrm{d}v^{\text{in}} + r^2 a(T)^2 \left( \mathrm{d} \theta^2 + \sin ^2 \theta \mathrm{d} \varphi^2 \right) \, . 
\end{equation}
Next, we incorporate the following double null coordinates for the exterior
\begin{equation} \label{uvout}
\begin{aligned}
u^{\text{out}}(t, X) = t - \int^X_0 \frac{1}{F(X')} \mathrm{d} X' \, ,\\
v^{\text{out}}(t, X) = t + \int^X_0 \frac{1}{F(X')} \mathrm{d} X' \, ,\\
\end{aligned}
\end{equation}
so that we have
\begin{equation} \label{metricoutdoublenull}
    \mathrm{d}s^2_+ = -F(X) \mathrm{d}u^{\text{out}}\mathrm{d}v^{\text{out}} +  X^2 \left( \mathrm{d} \theta^2 + \sin ^2 \theta \mathrm{d} \varphi^2 \right) \, .
\end{equation}
In the above, $F(X)$ is explicitly given by \eqref{FX}. With the pairs of double null coordinates established, we now solve for the parametrized trajectories of the boundary surface, here the surface of the collapsing dust ball $\Sigma$. It is convenient to choose $\dutchcal{s} \rightarrow T$ as the parameter at $\Omega \rightarrow \Sigma$. The trajectory w.r.t. coordinates \eqref{uvin} is simply given by a pair $(\tilde{u}^{\text{in}}(T, r = r_b), \tilde{v}^{\text{in}}(T, r=r_b))$. Subsequently, to obtain a trajectory w.r.t \eqref{uvout} one has to solve \eqref{gf} for $\dutchcal{t}(T)$. To get a solution we employ a similar strategy as in \cite{Bobula:2023kbo} (here with the initial condition $\dutchcal{t}'(0) = 0$) --- see the discussion around Eq. (2.29)-(2.31) there. Finally, the trajectory $(u^{\text{out}}(T), v^{\text{out}}(T))$ can be written as
\begin{equation} \label{uvouttraj}
\begin{aligned}
u^{\text{out}}(T) =u^{\text{out}}(t = \dutchcal{t}(T), X = R(T)) = \dutchcal{t}(T)- \int^{R(T)}_0 \frac{1}{F(X')} \mathrm{d} X' \, ,\\
v^{\text{out}}(T) =v^{\text{out}}(t = \dutchcal{t}(T), X = R(T)) = \dutchcal{t}(T) + \int^{R(T)}_0 \frac{1}{F(X')} \mathrm{d} X' \, .\\
\end{aligned}
\end{equation}
Note that the trajectories $u^{\mathrm{out}}(T)$ and $v^{\mathrm{out}}(T)$ are piecewise invertible on intervals $(-\infty, - T_3]$, $[-T_3, -T_2]$, $[-T_2,-T_1]$, $[-T_1, \infty]$ and $[-\infty,T_1]$, $[T_1, T_2]$, $[T_2,T_3]$, $[T_3, \infty)$ respectively.

Having the first step of the method of Section \ref{secMethod} accomplished, now we aim to collect points $(u^{\mathrm{in}}, v^{\mathrm{in}} )$ $(u^{\mathrm{out}}, v^{\mathrm{out}} )$ beyond $\Sigma$ --- our spacetime points. We begin with the exterior, that is, we look for $(u^{\mathrm{out}}, v^{\mathrm{out}} )$. Before we study the behaviour of an affine parameter on either $u^{\mathrm{out}} = const. $ or $v^{\mathrm{out}} = const.$ null geodesics, we point out some crucial properties. In particular, the null coordinates and the radial coordinate $X$ satisfy 
\begin{equation} \label{uvX}
\mathrm{d} v^{\mathrm{out}}-\mathrm{d} u^{\mathrm{out}}=\frac{2}{F(X)} \mathrm{d} X \, .
\end{equation}
Moreover, the $X$ is the affine parameter on null geodesics for the class of metrics of the form of \eqref{metricoutdoublenull} (see \cite{Bobula:2023kbo} for the proof\footnote{However, $X$ is not affine parameter on null geodesics when $F(X)=0$ (equivalently $X=const.$)}). Indeed, the equation \eqref{affineRelation} relating the null coordinate to the affine parameter reduces to $\mathrm{d} v^{\mathrm{out}}=\frac{2}{F(X)} \mathrm{d} X$ (for $u^{\mathrm{out}} =const.$ family of geodesics) or to $\mathrm{d} u^{\mathrm{out}}=\frac{-2}{F(X)} \mathrm{d} X$ (for $v^{\mathrm{out}} =const.$ family of geodesics). The integration of these will provide desired relations  $v^{\mathrm{out}} (X)$ and $u^{\mathrm{out}} (X)$ (equivalently $X(v^{\mathrm{out}})$ and $X(u^{\mathrm{out}})$). 

Next, we analyze null geodesics starting from $\Sigma$ propagating to the exterior. We can choose $T = T_{\mathrm{in}}$ at $\Sigma$ and then read off the initial $X=R(T_{\mathrm{in}})$ and  $u^{\mathrm{out}}(T_{\mathrm{in}}) $, $v^{\mathrm{out}}(T_{\mathrm{in}})$ (needed for the integration) from \eqref{Rsurf} and \eqref{uvouttraj} respectively. Let's consider $v^{\mathrm{out}} (T_{\mathrm{in}}) =const.$ geodesic departing from $\Sigma$. For the starting point $T_{\mathrm{in}}$, the function $u^{\mathrm{out}}(X) = \int_{R(T_{\mathrm{in}})}^X \frac{-2}{F(X)} \mathrm{d} X + u^{\mathrm{out}}(T_{\mathrm{in}})$ tells us how $u^{\mathrm{out}}$ varies with respect to the affine parameter. The range of $u^{\mathrm{out}}(X)$ is searched collection of $u^{\mathrm{out}}$ points, given $v^{\mathrm{out}} (T_{\mathrm{in}}) =const. $ geodesic. In this procedure, one has to qualitatively analyze all possible $v^{\mathrm{out}}(T_{\mathrm{in}}) =const. $, where $T_{\mathrm{in}} \in (-\infty, \infty)$. Similarly, for the $u^{\mathrm{out}} (T_{\mathrm{in}}) =const.$ geodesics, one analyses the function $v^{\mathrm{out}}(X) = \int_{R(T_{\mathrm{in}})}^X \frac{2}{F(X)} \mathrm{d} X + v^{\mathrm{out}}(T_{\mathrm{in}})$. 

The above procedure serves as a tool to probe the causal structure of the exterior. The supplement analysis is presented in Appendix \ref{appcollecting}. For the interior region, the situation is more straightforward. Namely, it is relatively easy to realize that \eqref{uvin} cover the whole interior\footnote{A congruence of timelike geodesic, each labelled by $r=const.$, $r \in [0,r_b]$ is complete since $T\in (-\infty, \infty)$ is the affine parameter on them. Indeed, the family of $r=const.$ ``tubes'' foliate the whole interior. } --- all of the point spacetime points are given by the pair $(\tilde{u}^{\text{in}}(T, r), \tilde{v}^{\text{in}}(T, r) )$, where $T \in (-\infty, \infty)$ and $r \in [0, r_b]$.

The next step is to extend a coordinate system. We extend the interior coordinates \eqref{uvin} to cover also the exterior. We proceed as follows: 
inverted versions of \eqref{uvouttraj}, namely $T(u^{\text{out}})$ and $T(v^{\text{out}})$, can be plugged to the pair $(\tilde{u}^{\text{in}}(T, r = r_b), \tilde{v}^{\text{in}}(T, r=r_b))$ to yield the desired extended coordinate system covering the exterior
\begin{equation} \label{extin}
    \left\{  
    \begin{aligned}
    \tilde{u}^{\text{in}}(u^{\text{out}}) := \tau(T(u^{\text{out}})) - r_b  \, ,\\
    \tilde{v}^{\text{in}} (v^{\text{out}}) := \tau(T(v^{\text{out}})) + r_b \, . \\
    \end{aligned}
    \right.
\end{equation}
Note that the ranges of \eqref{extin} are already compact due to the compactness of the conformal time $\tau$. As foreshadowed in Section \ref{secMethod}, the region covered by the extended coordinate system \eqref{extin}, due to the construction, is limited to the region enclosed by the first null ray leaving and the last one reaching $\Sigma$ -- see Figure \ref{uklady} for the general discussion. However, we aim to extend \eqref{extin} to cover spacetime regions beyond that. To get the extension of $(\tilde{u}^{\text{in}}(u^{\text{out}}), \tilde{v}^{\text{in}}(v^{\text{out}}))$ we proceed with simple polynomial extrapolation of these coordinates evaluated at $\Sigma$. 

\begin{figure}[h] 
    \centering
    \includegraphics[width=9cm]{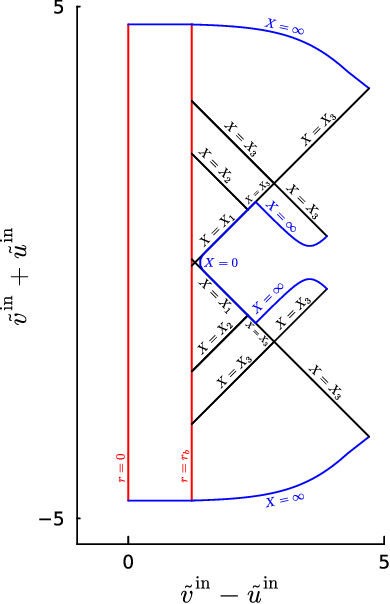}
    \caption{Conformal diagram of quantum dust ball collapse with cosmological constant according to extended coordinates \eqref{uvin}, \eqref{extin}. The collapsing surface of dust ball, indicated by the vertical line $r=r_b$, crosses 3 pairs of horizons $X_1$, $X_2$ and $X_3$ initially in the black-hole-like region and then in white-hole-like region. The interior of the dust ball is represented by the ``rectangle'' in the LHS side of the diagram. The collapse starts and ends at the surfaces $X=\infty$. The endpoints of the line representing timelike singularity, $X=0$, are at the points of intersection of horizons $X=X_2$ with surfaces of $X=\infty$ (not visible in the plot). The causal structure is the same as in Figures \ref{krzywy} and \ref{confout}. $G=c=\hbar=1$, and $\dutchcal{v}_{\mathrm{in}} = 10 $, $C=5 \times 10^{-5}$ (equivalently $M_{\mathrm{eff}}=13.88$, $\Lambda = 1.35 \times 10^{-5}$)} 
     \label{prosty}
\end{figure}

We are now ready to construct the conformal diagram. We aim to extract a ``real'' diagram, that is, we will numerically generate it on a computer. For that purpose, libraries of \textsc{Julia} programming language will be used. We present the example of such a diagram in Figure \ref{prosty}. To plot the interior, we utilize \eqref{uvin}. The surfaces $r=0$ and $r=r_b$ are simply given parametrically by $(\tilde{u}^{\text{in}}(T, r=0), \tilde{v}^{\text{in}}(T, r=0))$ and $(\tilde{u}^{\text{in}}(T, r=r_b), \tilde{v}^{\text{in}}(T, r=r_b))$ respectively. Analogously, surfaces $T \rightarrow \infty$ and $T \rightarrow - \infty$, parametrized with $r$, are given by $(\tilde{u}^{\text{in}}(T \rightarrow \infty, r), \tilde{v}^{\text{in}}(T \rightarrow \infty, r=0))$ and $(\tilde{u}^{\text{in}}(T \rightarrow -\infty, r), \tilde{v}^{\text{in}}(T \rightarrow -\infty, r))$. These four surfaces constitute a ``rectangle'' on the left side in Figure \ref{prosty}. To plot the exterior, we use \eqref{extin} and the analysis performed in Appendix \ref{appcollecting}. The $u^{\mathrm{out}}=const.$ horizons are given by $(u^{\mathrm{out}}(X)= const.,v^{\mathrm{out}}(T) )$, where $X\in \{X_1, X_2, X_3\}$ and $T$ is the corresponding time interval given at $\Sigma$ -- see Appendix \ref{appcollecting}. Similarly, the $v^{\mathrm{out}}=const.$ horizons are given by $(u^{\mathrm{out}}(T),v^{\mathrm{out}}(X) = const. )$. The next step is to simply plug the above pairs of points to \eqref{extin}, however, we have to remember the proper choice of the inverses $T(u^{\text{out}})$ and $T(v^{\text{out}})$. Indeed, plotting horizons simplifies to the determination of $T$ intervals, as $T(u^{\text{out}}(T))=T$ and $T(v^{\text{out}}(T))=T$. For example, $X=X_2$ horizon evolving towards the future (see Figure \ref{prosty}) is given by $(-T_2, T_1)$ plugged to $\tau(T) + r_b$ and $T=-T_2$ plugged to $\tau(T) - r_b$. Besides horizons, we also plot surfaces $X=0$ and $X \rightarrow \infty$. To do so, first, we can find the endpoints in the way presented in Appendix \ref{appcollecting}. For example, to plot the surface $X\rightarrow \infty$ in the bottom of Figure \ref{prosty} we collect $T_{\mathrm{in}}$ yielding $v^{\text{out}}(T_{\mathrm{in}})=const.$ geodesic and the corresponding $u^{\mathrm{out}}(X\rightarrow \infty)$. Alternatively, one has to specify which exactly maps of $T(u^{\mathrm{out}})$ and $T(v^{\mathrm{out}})$ are plugged to \eqref{extin} (remember about the piecewise invertibility), and then, the surfaces are simply given by respective pairs $(\tilde{u}^{\text{in}}(u^{\text{out}}(t, X=0)), v^{\text{out}}(t, X=0)) )$ and $(\tilde{u}^{\text{in}}(u^{\text{out}}(t, X\rightarrow\infty)), v^{\text{out}}(t, X\rightarrow \infty)) )$, where the parameter $t \in (-\infty, \infty)$. We note that horizons $X=X_3$ in Figure \ref{prosty} connected to the surfaces $X\rightarrow\infty$ cannot be crossed by geodesics departing from $\Sigma$, however, one can utilize the zig-zagging described in Section \ref{secMethod} in order to find the coordinate points needed for the plot. We also stress that the surface $X=0$ represents a timelike singularity -- the Kretschmann scalar diverges in there.

The derived conformal diagram in Figure \ref{prosty} depicts a collapsing dust ball forming a black hole evolving into a white hole. The collapsing matter crosses 3 pairs of horizons: $X_1, X_2$ and $X_3$. The causal structure in the exterior is the same as in the case of Reissner-Nordstr\"om-de Sitter solution to Einstein Field Equations -- see Figure 2 in \cite{Laue:1977zz}.  The diagram does not cover the entire spacetime, because it could be extended also beyond $X=X_3$ horizons in the right-hand side of Figure \ref{prosty} (see also Appendix \ref{appcollecting}). There, one could extend it to have infinitely many periodic copies of the exterior part of itself like in Figure 2 of \cite{Laue:1977zz}.

\begin{figure}[h] 
    \centering
    \includegraphics[width=9cm]{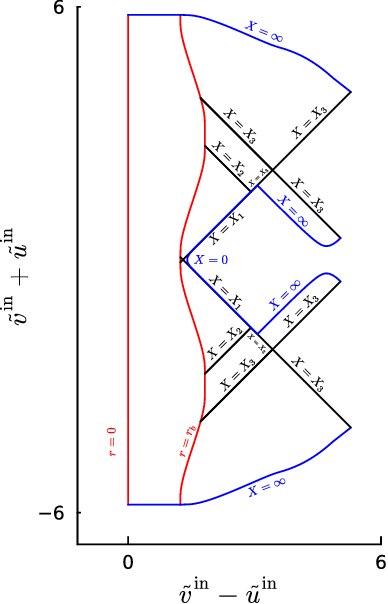}
    \caption{The conformal diagram of the dust collapse according to modified extended coordinates \eqref{uvinf}, \eqref{extinf}. The collapsing surface of dust ball, indicated by the curve $r=r_b$, crosses 3 pairs of horizons $X_1$, $X_2$ and $X_3$ initially in the black-hole-like region and then in white-hole-like region. The causal structure is completely the same as in Figures \ref{prosty} and \ref{confout}. The proposed function $f$ (explicitely given in Appendix \ref{appendixf}) ``enlarged'' the regions between horizons $X_1$ and $X_2$, as expected.  $G=c=\hbar=1$, and $\dutchcal{v}_{\mathrm{in}} = 10 $, $C=5 \times 10^{-5}$ (equivalently $M_{\mathrm{eff}}=13.88$, $\Lambda = 1.35 \times 10^{-5}$). So-called \textit{construction parameter} for the exterior coordinates is taken to be $s_0=0.01$ (see Section 5 of \cite{Schindler:2018wbx}). } 
     \label{krzywy}
\end{figure}

Furthermore, we show how to reshape the diagram in Figure \ref{prosty} to adjust it for our preferences. In particular, we want to ``enlarge'' the plotted region between horizons $X_1$ and $X_2$. To achieve so, we propose a following coordinate transformation of \eqref{uvin}
\begin{equation}  
\begin{aligned} \label{uvinf}
\tilde{u}^{\text{in}}(T, r) = f(\tau (T) - r) \, ,\\
\tilde{v}^{\text{in}}(T, r) = f(\tau (T) + r) \, ,\\
\end{aligned}
\end{equation}
and of \eqref{extin} 
\begin{equation} \label{extinf}
    \begin{aligned}
    \tilde{u}^{\text{in}}(u^{\text{out}}) = f(\tau(T(u^{\text{out}})) - r_b)  \, ,\\
    \tilde{v}^{\text{in}} (v^{\text{out}}) = f(\tau(T(v^{\text{out}})) + r_b) \, , \\
    \end{aligned}
\end{equation}
where $f$ is yet-to-be-proposed function. We want to put constraints on $f$ depending on how it should transform the diagram. Particularly, we aim to keep surfaces $(\tilde{u}^{\text{in}}(T \rightarrow \infty, r), \tilde{v}^{\text{in}}(T \rightarrow \infty, r))$ and $(\tilde{u}^{\text{in}}(T \rightarrow -\infty, r), \tilde{v}^{\text{in}}(T \rightarrow -\infty, r))$ as surfaces of constant $\tilde{v}^{\text{in}} + \tilde{u}^{\text{in}}$. That is, we have $f(\tau^{+} + r ) + f(\tau^{+} - r ) = const.$ and $f(\tau^{-} + r ) + f(\tau^{-} - r ) = const.$ respectively, where $\tau^{+}:= \lim_{T \rightarrow \infty } \tau(T) $ and $\tau^{-}:= \lim_{T \rightarrow -\infty } \tau(T) $. Equivalently, we can write these conditions as $f'(\tau^{+} + r ) - f'(\tau^{+} - r ) = 0$ and $f'(\tau^{-} + r ) - f'(\tau^{-} - r ) = 0$. It means that the function $f$ should be symmetric with respect to both $\tau^{-}$ and $\tau^{+}$ vertical lines. Subsequently, $f$ has to be monotonic -- we impose $f'>0$ for all of the domain. Finally, to acquire abovementioned enlargement, the values of $f'$ should be visibly bigger in the intervals $(-\tau^{(2)}, -\tau^{(1)})$ and $(\tau^{(1)}, \tau^{(2)})$ when compared to the other values in the domain $[\tau^{-}, \tau^{+}]$, where $\tau^{(1)}$ is the value of the conformal time evaluated at the point of intersection of $r=0$ with null geodesic propagated to the future from the point $T_1$ at $\Sigma$ (analogously for $\tau^{(2)}$). We construct $f$ explicitely in the Appendix \ref{appendixf}. Ultimately, with the derived $f$, we present the modified conformal diagram in Figure \ref{krzywy}.

\subsection{Extension of the exterior coordinates}
\begin{figure}[h] 
    \centering
    \includegraphics[width=13cm]{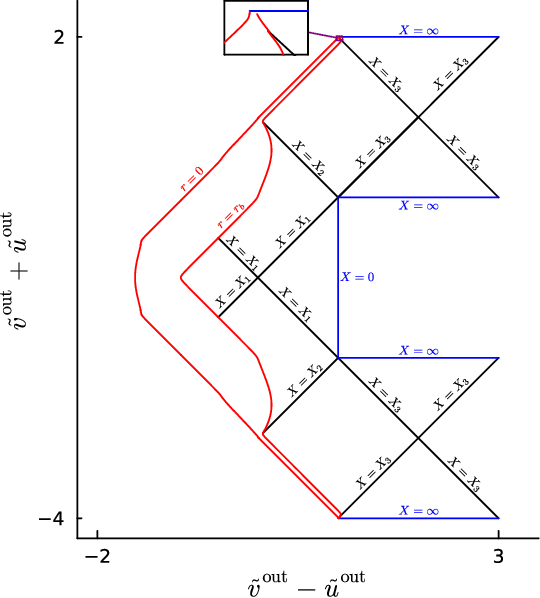}
    \caption{The conformal diagram of the dust collapse according to extended coordinates \eqref{algo}, \eqref{outtoin}. The collapsing surface of dust ball, indicated by the curve $r=r_b$, crosses 3 pairs of horizons $X_1$, $X_2$ and $X_3$ initially in the black-hole-like region and then in white-hole-like region. The causal structure is completely the same as in Figures \ref{prosty}, \ref{krzywy}. $G=c=\hbar=1$, and $\dutchcal{v}_{\mathrm{in}} = 10 $, $C=5 \times 10^{-5}$ (equivalently $M_{\mathrm{eff}}=13.88$, $\Lambda = 1.35 \times 10^{-5}$). \emph{A parameter of the construction} is taken to be $s=0.01$ (see Chapter 5.2 of \cite{Schindler:2018wbx} for the definition). } 
     \label{confout}
\end{figure}

As promised before, we also present an example of how the method (Section \ref{secMethod}) works when one initially starts with coordinates describing the exterior of the collapsing dust ball (the outcome is depicted in Figure \ref{confout}). So far, we have rewritten the exterior line element \eqref{metricoutdoublenull} with the double null coordinates \eqref{uvout} and also we found parametrized trajectory of the boundary surface \eqref{uvouttraj}. However, the coordinates \eqref{uvout} neither have compact ranges nor cover all of the spacetime regions of our interest within single coordinate chart. To cope with that, we utilize an algorithm existent in the literature \cite{Schindler:2018wbx} appropriate for the removal of these obstacles via suitable coordinate transformation. More specifically, the algorithm not only compactifies double null coordinates arising for the class of metrics of the form of \eqref{metricoutdoublenull} ($F(X)$ can have any form), but also allows to cover arbitrarily large region (described by the metric of the form of \ref{metricoutdoublenull}) within single coordinate chart. Following Section V of \cite{Schindler:2018wbx} we write coordinate transformation
\begin{equation} \label{algo}
        \begin{aligned}
    \tilde{u}^{\text{out}}(u^{\text{out}}) = \arctan \left(\epsilon_{u^{\text{out}}} h(u^{\text{out}} / 2) \right)/\pi + \tilde{c}_{u^{\text{out}}} /\pi  \, ,\\
    \tilde{v}^{\text{out}} (v^{\text{out}}) = \arctan \left(\epsilon_{v^{\text{out}}} h(v^{\text{out}} / 2) \right)/\pi + \tilde{c}_{v^{\text{out}}} /\pi \, , \\
    \end{aligned}
\end{equation}
where $h$ is so-called \textit{pre-squishing function} (see Section 5.2 of \cite{Schindler:2018wbx} for explicit form and more details), $\epsilon_{u^{\text{out}}}$ and $\epsilon_{v^{\text{out}}} $ take values in $\{-1, 1\}$ determining \textit{orientation} of a \textit{block} (a basic building component\footnote{For example a conformal diagram of Schwarzschild spacetime consists of 4 blocks: black hole interior, white hole interior, a universe and a parallel universe.} of a conformal diagram, see Section 4.4 of \cite{Schindler:2018wbx} for a rigorous definition), constants $\tilde{c}_{u^{\text{out}}}$ and $\tilde{c}_{v^{\text{out}}}$ are responsible the location of each block. That is, in each block, constants present in \eqref{algo} are not the same. Having that setup, it is straightforward to follow the rules of \cite{Schindler:2018wbx} to plot the exterior parts (of our interest) of the conformal diagram. However, since we are interested in covering also the interior with \eqref{algo} we need to extend the coordinates. We note that the parametrized trajectory of the surface of the collapsing dust ball in new coordinates \eqref{algo} is the pair
\begin{equation} \label{outtraje}
        \begin{aligned}
    \tilde{u}^{\text{out}} (T)  = \arctan \left(\epsilon_{u^{\text{out}}} h(u^{\text{out}}(T)/ 2) \right)/\pi + \tilde{c}_{u^{\text{out}}} /\pi  \, ,\\
    \tilde{v}^{\text{out}} (T) = \arctan \left(\epsilon_{v^{\text{out}}} h(v^{\text{out}}(T) / 2) \right)/\pi + \tilde{c}_{v^{\text{out}}} /\pi \, . \\
    \end{aligned}
\end{equation}
where \eqref{uvouttraj} was utilized. One has to be careful about the switching of the constants in \eqref{algo} while the surface of the collapsing dust ball cross from one block to another. The extended coordinates are
\begin{equation} \label{outtoin}
    \left\{  
    \begin{aligned}
    \tilde{u}^{\text{out}}(\tilde{u}^{\text{in}}) := \arctan \left(\epsilon_{u^{\text{out}}} h(u^{\text{out}}(T(\tilde{u}^{\text{in}}))  / 2) \right)/\pi + \tilde{c}_{u^{\text{out}}} /\pi  \, ,\\
    \tilde{v}^{\text{out}} (\tilde{v}^{\text{in}}) := \arctan \left(\epsilon_{v^{\text{out}}} h(v^{\text{out}}(T(\tilde{v}^{\text{in}}))   / 2) \right)/\pi + \tilde{c}_{v^{\text{out}}} /\pi \, , \\
    \end{aligned}
    \right.
\end{equation}
where the pair $(\tilde{u}^{\text{in}}, \tilde{v}^{\text{in}})$ is given by \eqref{uvin} and $T(\tilde{u}^{\text{in}})$, $T(\tilde{v}^{\text{in}})$ are inverses of $\tilde{u}^{\text{in}}(T, r = r_b) $ and $\tilde{v}^{\text{in}}(T, r=r_b)$ respectively.

The above setup is sufficient to generate the corresponding conformal diagram -- Figure \ref{confout}. The boundary surface $\Sigma$ is given by \eqref{outtraje}. To plot the exterior with \eqref{algo} we straightforwardly follow the algorithm of Section V of \cite{Schindler:2018wbx}. Although, as before, the diagram could have possessed infinitely many copies of itself in the exterior part, we restrict the plot to cover the same region as in Figures \ref{prosty}, \ref{krzywy}. Subsequently, the surface $r=0$ in the interior is given by $(\tilde{u}^{\text{out}}(\tilde{u}^{\text{in}} (T, r=0)) , \tilde{v}^{\text{out}}(\tilde{v}^{\text{in}} (T, r=0)) )$ with $T \in (-\infty, \infty)$. Of course, we also plot interior surfaces $T\rightarrow \pm \infty$ given by $(\tilde{u}^{\text{out}}(\tilde{u}^{\text{in}} (T \rightarrow \pm \infty, r)) , \tilde{v}^{\text{out}}(\tilde{v}^{\text{in}} (T \rightarrow \pm \infty, r)) )$ with $r \in [0, r_b]$.

\section{Conclusions}
\label{secConclusions}

We have studied the process of quantum-corrected Oppenheimer-Snyder dust ball collapse within the otherwise vacuum spacetime with positive cosmological constant. The effective geometry of the Friedmann-Robertson-Walker interior was determined by Loop Quantum Cosmology where the quantum Hamiltonian was supplemented with the term reflecting the presence of the cosmological constant. Indeed, the obtained semiclassical interior geometry was a result of solving effective equations of motion for expectation values of quantum observables. The resulting trajectories of these (expectation values of) observables delivered the metric components of our interior dust universe. The exterior geometry was not a subject of loop quantization. Instead, it was determined thanks to the analyzed junction conditions formalism. That is, by imposing i) a general form of spherically symmetric metric for the exterior, ii) the existence of the timelike Killing vector in there, we were able to uniquely determine the exterior metric solely by Israel-Darmois junction conditions at the boundary --- the surface of the collapsing dust ball. Similar techniques were recently studied in \cite{Bobula:2023kbo, Bobula:2024jlh, Lewandowski:2022zce, Bonanno:2023rzk}, however, unlike in these works, we also explored the formalism when the exterior could be time-dependent (absence of the timelike Killing vector). Accordingly, from the junction conditions formalism we learn that given a certain class of dust FRW universes, there are compatible (unique) black hole spacetimes and vice versa given black hole spacetimes there are unique dust universes.

In our model of consideration, the causal structure of the collapsing dust ball resembles Reissner-Nordstr\"om-de Sitter spacetime. The Penrose-Carter diagrams are displayed in Figures \ref{prosty}, \ref{krzywy} and \ref{confout} (note that all of them represent the same causal structure). The surface of the collapsing dust ball crosses 3 pairs of horizons. First, it crosses 3 black-hole-like horizons, and subsequently bounces and crosses 3 white-hole-like horizons. One aspect of the motivation for this work was to determine whether a worrisome feature of models presented in \cite{Bobula:2023kbo, Lewandowski:2022zce}, namely, the existence of the timelike singularity, persists in the dust collapse scenario with the cosmological constant. The answer turned out to be positive, however, we emphasize that the exterior geometry does not directly come from the LQC dynamics but rather is a result of the employed junction conditions formalism. Perhaps to obtain a more robust picture, one should loop quantize both componential geometries, that is, the interior and exterior of the dust ball. We also observe that the inner horizon instabilities (see \cite{Poisson:2009pwt} for a general discussion) might also plague the model presented in this work (this issue also concerns the models without cosmological constant \cite{Bobula:2023kbo, Lewandowski:2022zce}). Nevertheless, this problem might be resolved by considering a more realistic scenario accounting for the backreaction of Hawking radiation. Recent studies, on more general grounds, support this possibility \cite{Bonanno:2022jjp}.

As a part of this work, we refined existing methods for (numerical) extraction of Penrose-Carter (conformal) diagrams. In Sec. \ref{secMethod} we proposed methods for: i) identifying the positions of the spacetime points on the would-be-generated diagram and ii) extending the coordinate system from the first componential spacetime to cover also the second componential one (within the single coordinate chart). In Sec. \ref{secDIAG}, the developed tools were applied to the studied model of dust ball collapse in the presence of cosmological constant. The diagrams in Figures \ref{prosty}, \ref{krzywy} and \ref{confout} seemingly "look different", but they represent the same causal structure. The presented tools are useful for the rigorous extraction of conformal diagrams. To avoid confusion and incorrect results, such a rigorous treatment is necessary in several situations where the complicatedness of the studied problem is significant.

\section*{ACKNOWLEDGEMENTS}
This work was supported in part by the Polish National Center for Science (Narodowe Centrum Nauki –
NCN) grant OPUS 2020/37/B/ST2/03604.

\appendix

\section{Roots of $F(X)$} 
\label{approots}

\begin{figure} 
    \centering
    \includegraphics[width=10cm]{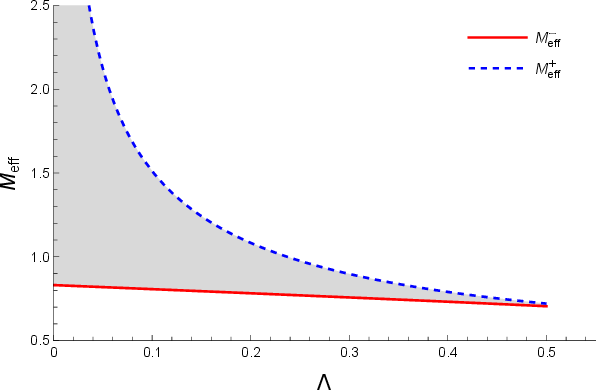}
    \caption{Roots of \eqref{disc} as functions of the cosmological constant, these are $M_{\mathrm{eff}}^{+}(\Lambda)$ (solid line) and $M_{\mathrm{eff}}^{-}(\Lambda)$ (dashed line). The domain of the plot is $\Lambda \in [0, \frac{(2 \sqrt{2} -3) A \hbar}{8 \alpha_o \rho_c}]$. At the point $\Lambda =\frac{(2 \sqrt{2} -3) A \hbar}{8 \alpha_o \rho_c} $, $M_{\mathrm{eff}}^{+}(\Lambda)$ and $M_{\mathrm{eff}}^{-}(\Lambda)$ intersect. The greyed out region indicate the case $M_{\mathrm{eff}}^-(\Lambda) < M_{\mathrm{eff}} < M_{\mathrm{eff}}^+(\Lambda) $, when $F(X)$ has 3 roots. $G=c=\hbar=1$.}
    \label{appRYS}
\end{figure}

While the equation \eqref{FX} is which roots identify the horizons, as obtained via semiclassical methods, is reliable only when the spacetime is highly semiclassical, one can in principle probe it in its whole domain of definition.
Here, we analyze a number of possible roots of this equation depending on values of the mass of the dust ball and the cosmological constant. Due to the complexity of finding the explicit solutions of $F(X)=0$ we aim to study a discriminant of the polynomial $P(X):= F(X) X^4$ instead. We have
\begin{equation} \label{disc}
    \begin{split}
\mathrm{Disc}_{P(X)} = & \frac{5184}{(A-2  C)^{10}}  C  G^6 M_{\mathrm{eff}}^6 (A-C) \left(144 A^6 C G^4 M_{\mathrm{eff}}^4
   (A-C)- \right.\\
  & \left.3 G^2 M_{\mathrm{eff}}^2 (A-2 C)^2 \left(A^4+72 A^3 C-504 A^2
   C^2+864 A C^3-432 C^4\right)+4 (A-2
   C)^4\right) \, .
    \end{split}
\end{equation}
Subsequently, we solve $\mathrm{Disc}_{P(X)}(M_{\mathrm{eff}})=0 $ and we find that there are two positive roots of \eqref{disc}
\begin{equation} \label{M+-}
\begin{aligned} 
       M^\pm_{\mathrm{eff}} = 
   \frac{1} {4 \sqrt{6}} \bigg( \frac{1}{A^6 C G^4
   (A-C)}  \Big( G^2 (A-2 C)^2 \left(A^4+72 A^3 C-504 A^2
   C^2+864 A C^3- 432 C^4\right)   \\
   \pm \sqrt{G^4 (A-2 C)^6 
   \left(A^2-36 A C+36 C^2\right)^3} \Big) \bigg)^{1/2} 
\end{aligned}
\end{equation}
We observe that the above $M^\pm_{\mathrm{eff}}$ are real as long as $A^2 - 36 A C + 36 C^2 \geq 0$. We restrict to $C \in [0, \frac{A}{6} ( 3- 2 \sqrt{2})]$ (equivalently $\Lambda \in [0, \frac{(2 \sqrt{2} -3) A \hbar}{8 \alpha_o \rho_c}]$, remember that $C:=  \frac{3\hbar}{4\rho_c\alpha_o} \Lambda$), a branch where the quadratic inequality is satisfied, since we are interested in scenarios where the cosmological constant is rather \emph{small} and comparable to observational values. In other words, we aim to study the case when the dust ball collapse is ``slightly perturbed'' by the presence of the cosmological constant, consequently, in the limit $\Lambda \rightarrow 0$ we should retrieve the model presented in \cite{Bobula:2023kbo}. We observe that the roots \eqref{M+-} written as functions of the cosmological constant, namely $M^\pm_{\mathrm{eff}}(\Lambda)$, indicate \emph{critical} lines of \emph{transition} between numbers of possible roots of $F(X)$. More precisely, we end up with the following possibilities (visualized in the Figure \ref{appRYS}):
\begin{itemize}
    \item $M_{\mathrm{eff}} (\Lambda) > M_{\mathrm{eff}}^+ (\Lambda)$. $F(X)$ has one positive root. 
    \item  $M_{\mathrm{eff}} (\Lambda) = M_{\mathrm{eff}}^+ (\Lambda)$. First extreme case when $F(X)$ has one positive root exactly at the point where $F'(X)=0$. 
    \item $M_{\mathrm{eff}}^-(\Lambda) < M_{\mathrm{eff}} < M_{\mathrm{eff}}^+(\Lambda) $. $F(X)$ has 3 roots. This case is depicted as greyed out region in Figure \ref{appRYS}. 
    \item $M_{\mathrm{eff}} (\Lambda) = M_{\mathrm{eff}}^- (\Lambda)$. Second extreme case when $F(X)$ has one positive root exactly at the point where $F'(X)=0$.
    \item $M_{\mathrm{eff}} (\Lambda) < M_{\mathrm{eff}}^- (\Lambda)$. $F(X)$ has one positive root.
\end{itemize}
We note that for the observable value of the cosmological constant $\Lambda_{\mathrm{obs}} \approx 3 \times 10^{-122} m^2_{\mathrm{Pl}}$ (``Planck 2018 results'' \cite{Planck:2018vyg}) we have $M_{\mathrm{eff}}^+(\Lambda_{\mathrm{obs}}) \approx 4 \times 10^{59} \, m_{\mathrm{Pl}} \approx 4 \times 10^{21}$ \(M_\odot\) and $M_{\mathrm{eff}}^-(\Lambda_{\mathrm{obs}}) \approx 0.83 \,  m_{\mathrm{Pl}}$. Thus, in order to be close to the astrophysical scenarios of gravitational collapse, we restrict to the case $M_{\mathrm{eff}}^-(\Lambda) < M_{\mathrm{eff}} < M_{\mathrm{eff}}^+(\Lambda) $ where $F(X)$ has 3 roots. We also stress that $ \lim_{\Lambda \rightarrow 0} M_{\mathrm{eff}}^-(\Lambda) = 16 \alpha_o r_b^3 /\left(9 \sqrt{3} G^2 \hbar\right) \approx \, 0.83 \, m_{\mathrm{Pl}}$ -- in that limit the extreme mass is the same as in the model without cosmological constant \cite{Bobula:2023kbo}, as expected. 

\section{Probing the causal structure of the exterior} \label{appcollecting}

One of the crucial steps in constructing the Penrose-Carter diagram is the analysis of null geodesics. Here we present in more detail the procedure applied in the main body of the paper to the exterior region. The main tool are the functions relating null coordinates to affine parameter. They take the form $u^{\mathrm{out}}(X) = \int_{R(T_{\mathrm{in}})}^X \frac{-2}{F(X)} \mathrm{d} X + u^{\mathrm{out}}(T_{\mathrm{in}})$ for $v^{\mathrm{out}} = const.$ geodesics and  $v^{\mathrm{out}}(X) = \int_{R(T_{\mathrm{in}})}^X \frac{2}{F(X)} \mathrm{d} X + v^{\mathrm{out}}(T_{\mathrm{in}})$ for $u^{\mathrm{out}} = const.$ geodesics respectively. 

Let us first consider $v^{\mathrm{out}} =const.$ geodesics starting at $\Sigma$ at points labeled by $T=T_{\mathrm{in}}$ and then propagating towards the exterior:
\begin{itemize}
   \item $ T_{\mathrm{in}} \in (-\infty, -T_3)$. $X$ increases, $u^{\mathrm{out}}(X)$ has a compact range, and $u^{\mathrm{out}}(X \rightarrow \infty) = const.(T_{\mathrm{in}})$.
    \item $ T_{\mathrm{in}} \in (-T_3, -T_2)$. $X$ increases, $u^{\mathrm{out}}(X)$ diverges at $X_3$ and $u^{\mathrm{out}}(X \rightarrow \infty) = const.(T_{\mathrm{in}})$.
    \item $ T_{\mathrm{in}} \in (-T_2, -T_1)$. $X$ increases, $u^{\mathrm{out}}(X)$ diverges at $X_2$ and $X_3$ (the order does matter). $u^{\mathrm{out}}(X \rightarrow \infty) = const.(T_{\mathrm{in}})$.
    \item $ T_{\mathrm{in}} \in (-T_1, T_1)$.  $X$ increases, $u^{\mathrm{out}}(X)$ diverges at $X_1$, $X_2$ and $X_3$. $u^{\mathrm{out}}(X \rightarrow \infty) = const.(T_{\mathrm{in}})$. 
    \item $ T_{\mathrm{in}} \in (T_1, T_2)$. $X$ decreases, $u^{\mathrm{out}}(X)$ diverges at $X_1$. $u^{\mathrm{out}}(X \rightarrow 0) = const.(T_{\mathrm{in}})$.
    \item $ T_{\mathrm{in}} \in (T_2, T_3)$. $X$ increases, $u^{\mathrm{out}}(X)$ diverges at $X_3$. $u^{\mathrm{out}}(X \rightarrow \infty) = const.(T_{\mathrm{in}})$.
    \item $ T_{\mathrm{in}} \in (T_3, \infty)$. $X$ decreases, $u^{\mathrm{out}}(X)$ diverges at $X_3$, $X_2$ and $X_1$. $u^{\mathrm{out}}(X \rightarrow 0) = const.(T_{\mathrm{in}})$.
\end{itemize}
We emphasize that $X \rightarrow \infty$ points indicate spacetime endpoints (infinite value of the affine parameter). Also, for $X\rightarrow 0$ the Kretschmann scalar corresponding to the metric \eqref{metricoutdoublenull} is divergent -- these points represent the timelike singularity. The analysis for the last time interval, $ T_{\mathrm{in}} \in (T_3, \infty)$ suggest that the exterior region possesses infinitely many copies of itself (typical feature for Reissner-Nordstr\"om-like spacetimes). The goedesics $ T_{\mathrm{in}} \in (T_1, T_2)$ and $ T_{\mathrm{in}} \in (T_3, \infty)$  hit causally distinct timelike singularities. One can later confirm it by plotting the collected $u^{\mathrm{out}}$ on the conformal diagram. However, for the generated conformal diagrams presented in this work, we do not plot the spacetime beyond $X_3$ in the case of $ T_{\mathrm{in}} \in (T_3, \infty)$. Besides, the points where $u^{\mathrm{out}}(X)$ diverges at $X \in \{X_1, X_2, X_3\}$ represent horizons -- the corresponding $u^{\mathrm{out}}(X) = const.$ geodesic, from \eqref{uvX}, gives $\frac{\mathrm{d}X}{\mathrm{d}v^{\mathrm{out}}} = 0 \rightarrow X(v^{\mathrm{out}}) =const.$ (keep in mind that $X \in \{X_1, X_2, X_3\}$ are the roots of $F(X)$). In other words, surfaces $X \in \{X_1, X_2, X_3\}$ are null. By the similar argument, at $v^{\mathrm{out}}=const.$ geodesics, we can conclude that for starting points $T_{\mathrm{in}} \in \{T_1,T_2, T_3 \}$ geodesics represent horizons.

One can repeat the above analysis for the $u^{\mathrm{out}}=const.$ family of geodesics. The outcome will be similar, but time-reversed $T\rightarrow -T$. For example, for $ T_{\mathrm{in}} \in (-\infty, -T_3)$, $X$ decreases, $v^{\mathrm{out}}(X)$ diverges at $X_3$, $X_2$ and $X_1$. $v^{\mathrm{out}}(X \rightarrow 0) = const.(T_{\mathrm{in}})$ and so on.

\section{Construction of the function $f$} \label{appendixf}

Here we provide the construction of an explicit form of the function $f$ satisfying properties listed in Section\ref{extenstionInteriorSEC}. Firstly, we define $\dutchcal{y} (x):= - \cos(\frac{\pi}{\tau^+} x) $. Let $y^{(1)}:= y( \tau^{(1)} ) $, $y^{(2)}:= y( \tau^{(2)} ) $, and also
\begin{equation}
    \begin{aligned}
        a^{(y)} := & -2/(y^{(1)} - y^{(2)} ) \, . \\
        b^{(y)} := & \frac{y^{(1)} + y^{(2)}}{y^{(1)} - y^{(2)}} \, . \\
    \end{aligned}
\end{equation}
Next, we define $z(\dutchcal{y}) := a^{(y)} \dutchcal{y} +b^{(y)} $. Subsequently, let
\begin{equation}
    \dutchcal{h} (z):= \begin{cases}
0 \quad \mathrm{for}  \quad |z| \leq 1 \, ,\\
a^{(y)} \exp \left( -\frac{1}{1-z^2} \right) \quad \mathrm{for} \quad |z| \geq 1 \, .
\end{cases}
\end{equation}
Finally we write $f'(x):= 1+\dutchcal{h}(z(\dutchcal{y}(x))) $, and $f(x) = \int^{x}_0 f'(x') \mathrm{d}x' $. 

\bibliography{sample}

\end{document}